\documentclass[12pt]{article}

\usepackage{amsmath,amssymb}
\usepackage{graphicx}
\usepackage{fullpage}
\usepackage{hyperref}

\usepackage{color}

\newtheorem{theorem}{Theorem}[section]
\newtheorem{lemma}[theorem]{Lemma}

\newcommand{\R}{\mathbb{R}}
\newcommand{\Z}{\mathbb{Z}}
\newcommand{\proof}{\noindent{\bf Proof:}~}
\newcommand{\qed}{\hfill{\bf QED}\vspace{3mm}}
\newcommand{\e}{\mathfrak{e}}
\renewcommand{\v}{\mathfrak{v}}
\newcommand{\cS}{\mathcal{S}}
\newcommand{\E}{\mathcal{E}}
\newcommand{\V}{\mathcal{V}}

\newcommand{\red}[1]{{#1}}

\begin{document}

\title{Designing heteroclinic and excitable networks in phase space using two populations of coupled cells}

\author{Peter Ashwin\\
Center for Systems, Dynamics and Control\\
University of Exeter\\
Exeter EX4 4QF, UK
\and 
Claire Postlethwaite\\
Department of Mathematics\\
University of Auckland\\
Auckland, New Zealand}

\maketitle

\begin{abstract}
We give a constructive method for realizing an arbitrary directed graph (with no one-cycles) as a heteroclinic or an excitable dynamic network in the phase space of a system of coupled cells of two types. In each case, the system is expressed as a system of first order differential equations. One of the cell types (the $p$-cells) interacts by mutual inhibition and classifies which vertex (state) we are currently close to, while the other cell type (the $y$-cells) excites the $p$-cells selectively and becomes active only when there is a transition between vertices. We exhibit open sets of parameter values such that these dynamical networks exist and demonstrate via numerical simulation that they can be attractors for suitably chosen parameters.
\end{abstract}

\section{Introduction}
\label{sec:intro}

Researchers in neuroscience often want to understand how the structure of the central nervous system of animals relates to the function of the system both in healthy and diseased individuals, and much effort has been spent trying to model coupled neurons as networks of nonlinearly interacting cells (see e.g. \cite{Izhekevich,Ermentrout}) and emergent dynamical properties of the network are clearly important for an understanding of neural function from basic signal processing to high-level cognition; for example \cite{Kopell2014} suggest that the ``Dynome'' of possible states of the network is just as critical as the physical ``Connectome'' of connections between neurons and neural assemblies.

In addition to the physical network of coupled cells, the possible transient dynamical states of the system may usefully be thought to form a network {\em in phase space} of some type. \red{A {\em heteroclinic network} is a finite set of dynamical states connected by trajectories (defined more precisely below). } This idea has been developed in a number of models inspired by neural systems \cite{aguiar_ashwin_dias_field_11,ashwin_borresen_04,wordsworth_ashwin_08,komarov_osipov_suykens_09}; see also winnerless competition dynamics \cite{Rabinovich2001,ashwin_lavric_2010}, stable heteroclinic channels \cite{bick_rabinovich_2010}, and networks of unstable attractors \cite{neves_timme_12}. \red{If there are no ``direct connections'' between states, there may be {\em excitable connections} where a certain amplitude of a on-off perturbation at a state is needed for a transition from one state to another, giving rise to an {\em excitable network} (also defined below) whose structure will typically depend on the amplitude of perturbation used.\footnote{\red{Excitable networks in this sense have been considered before (for example see \cite{AshOroWorTow07}), but should not be confused with networks of coupled excitable units (for example see \cite{Linder_etal_2004}), that may or may not have excitable networks in phase space, depending on the coupling.}}}

In a previous paper \cite{AshPos13}, we proposed two constructions to show how arbitrary graphs may be embedded or {\em realised} as a heteroclinic network in the sense that there is a one-to-one mapping between vertices and edges of the graph and dynamical states. That paper presents a ``simplex network'' and a ``cylinder network'' of coupled cells that allows one to realise any finite directed graph that is one-cycle free into phase space as a heteroclinic network. The vertices of the graph correspond to equilibria that are saddles in phase space and the edges of the graph correspond to connecting (heteroclinic) orbits in phase space. The constructions in~\cite{AshPos13} require several different cell types; Field \cite{Field14} has recently shown that it is possible to do this even if one restricts to just one cell type.

This paper aims to present an explicit construction to realize arbitrary graphs in phase space as either heteroclinic or excitable networks. \red{ Not only this, we identify a bifurcation from a heteroclinic network to an excitable network of the same topology on changing a single parameter in the governing equations. } In the former case the vertices are saddle equilibria and the connections are heteroclinic. In the latter case the vertices correspond to stable equilibria in the network that are sensitive to perturbations in directions corresponding to the edges in the graph. The network construction uses two cell types, where there is strong inhibition between cells of one type and strong excitation of this cell type by the second cell type, reminiscent of neuronal systems. 


We structure the paper as follows: after defining heteroclinic and excitable networks in phase space we give a simple motivating example (given in equation~(\ref{eq:C3system})) of a system that realises a cyclic graph between three nodes as either a heteroclinic or an excitable network, depending on values of the parameters in the equations. In Section~\ref{sec:model} we introduce an explicit description of a coupled cell model (given in equation~(\ref{eq:realiseode})) that is parameterised by a number of constants. \red{ The first main result, Theorem~\ref{thm:realisehet}, shows that an arbitrary finite directed network can be robustly realised as a heteroclinic network in the phase space of this coupled cell system. Similarly, the second main result, Theorem~\ref{thm:realiseexc}, shows that an arbitrary finite directed network can be robustly realised as an excitable network for amplitude $\delta$ in the phase space of this coupled cell system. The minimum amplitude is related to the distance of a parameter from a bifurcation point where the saddles in the heteroclinic network are stabilized. In particular, the $\delta$ may be made as small as desired by choosing parameters appropriately.}

We give some numerical examples in Section~\ref{sec:examples} that realise the Kirk--Silber network \cite{KS94} of competition between two cycles using (\ref{eq:realiseode}) in the two different ways outlined in Section~\ref{sec:model}. In the presence of noise, we note that typical trajectories explore either cycle in a random manner. In Section~\ref{sec:discuss} we discuss some implications of the study, including generalizations that give networks in phase space where there may be a mixture of heteroclinic and excitable connections, and where the excitable connections may have various thresholds.

\subsection{Heteroclinic and excitable networks in phase space}
\label{subsec:hetexcit}

Consider an ODE with phase space $x\in\R^d$, defined by
\begin{equation}
\label{eq:ode}
\frac{dx}{dt}=\dot{x}=f(x),
\end{equation}
and suppose that the flow generated by the solution of this ODE starting at $x_0$ is $x(t)=\phi_t(x_0)$. Let $B_{\delta}(x)$ denote the closed ball centered on $x$ with radius $\delta>0$. For $\xi$ an equilibrium of~\eqref{eq:ode} we define the stable and unstable sets $W^s(\xi)=\{y~:~|\phi_t(y)-\xi|\rightarrow 0\mbox{ as }t\rightarrow \infty\}$ and $W^u(\xi)=\{y~:~|\phi_t(y)-\xi|\rightarrow 0\mbox{ as }t\rightarrow -\infty\}$; these are manifolds if $\xi$ is hyperbolic. Typically we will consider the case that all equilibria are hyperbolic.

\red{Our definition of a heteroclinic network is substantially weaker than that given in most of the literature (see below for further details); }
we say a set $X\subset \R^d$ is a \red{{\em (weak) heteroclinic network (in phase space)} \footnote{\red{We refer to  a ``(weak) heteroclinic network (in phase space)'' simply as a ``heteroclinic network'' for the remainder of the paper.}} } if there is a set of equilibria $\{\xi_i\}_{i=1}^n$ such that
$$
X=X_{\mathrm{het}}(\{\xi_i\}):=\bigcup_{i,j=1}^{n} W^u(\xi_i)\cap W^s(\xi_j)
$$
and we say there is a {\em heteroclinic connection} from $\xi_i$ to $\xi_j$ whenever
$$
W^u(\xi_i)\cap W^s(\xi_j)\neq \emptyset.
$$
We assume there are no homoclinic connections, i.e.\ that $W^u(\xi_i)\cap W^s(\xi_i)=\{\xi_i\}$. Note that this definition of heteroclinic network is weaker than that used in most of the literature, e.g.~\cite{KPR13,AshPos13}, in the following ways: \red{ (a) we do not require any chain recurrence or even connectedness of the network; for example, we do not exclude the possibility that the system is of gradient type; (b) we do not require that the entire unstable set is contained in the network; (c) we do not require that the equilibria are hyperbolic, although in typical cases the equilibria of heteroclinic networks are saddles (if there are incoming and outgoing heteroclinic connections at that equilibria), and the equilibria of proper excitable networks are sinks.}

We say a set $X\subset \R^d$ is an {\em excitable network} (in phase space) for {\em amplitude} $\delta>0$ if there is a set of equilibria $\{\xi_i\}_{i=1}^n$ such that
$$
X=X_{\mathrm{exc}}(\{\xi_i\},\delta):=\bigcup_{i,j=1}^{n} \{\phi_t(x)~:~x\in B_{\delta}(\xi_i) \mbox{ and }t>0\}\cap W^s(\xi_j)
$$
In other words, an excitable network is the union of a number of equilibria and the set of trajectories within the stable manifolds of these equilibria that come within $\delta$ of other equilibria. We say there is an {\em excitable connection for amplitude} $\delta>0$ from $\xi_i$ to $\xi_j$ whenever
$$
B_{\delta}(\xi_i)\cap W^s(\xi_j)\neq \emptyset.
$$
\red{ We say an excitable connection from $\xi_i$ to $\xi_j$ has {\em threshold} $\delta_{th}(\xi_i,\xi_j)$ where
\begin{equation}
\delta_{th}(\xi_i,\xi_j): = \inf \{\delta>0~:~B_{\delta}(\xi_i)\cap W^s(\xi_j)\neq \emptyset\}.
\label{eq:deltath}
\end{equation}
An excitable network for amplitude $\delta$ is {\em proper} if all of its excitable connections have finite threshold, i.e. if there is a $\delta'$, with $\delta>\delta'>0$ such that there are no excitable connections for amplitude $\delta'$ from any $\xi_i$ to any $\xi_j$ within the network. In this terminology, a heteroclinic connection from $\xi_i$ to $\xi_j$ corresponds to there being an excitable connection with zero threshold.}

An excitable network $X$ is forwards (but not necessarily backwards) invariant ($\phi_t(X)\subset X$ for all $t>0$) while a heteroclinic network is both forwards and backwards invariant ($\phi_t(X)\subset X$ for all $t\in\R$). Observe also that any finite set of equilibria are connected in an all-to-all manner by taking excitable connections with a large enough amplitude.

Now consider a (finite) graph $\Gamma=(\V,\E)$ with $n_v$ vertices $\V=\{\v_1,\ldots,\v_{n_v}\}$ and $n_e$ directed edges $\E=\{\e_1,\ldots,\e_{n_e}\}$. We define $\alpha(k)$ and $\omega(k)$ so that $\e_k$ is the edge from $\v_{\alpha(k)}$ to $\v_{\omega(k)}$. We say $\Gamma$ is one-cycle free if $\alpha(k)\neq\omega(k)$ for all $k$ and we will assume henceforth that $\Gamma$ is one-cycle free. We say a heteroclinic network $X$ {\em realises} the graph $\Gamma$ if each vertex $\v_i$ of $\Gamma$ corresponds to an equilibrium $\xi_i$ in $X$, and there is an edge of $\Gamma$ from $\v_i$ to $\v_j$ if and only if there is a connection from $\xi_i$ to $\xi_j$ in $X$. We say an excitable network $X$  for amplitude $\delta$ {\em realises} the graph $\Gamma$ if each vertex $\v_i$ in $\Gamma$ corresponds to an equilibrium $\xi_i$ in $X$ and there is an edge in $\Gamma$ from $\v_i$ to $\v_j$ if and only if there is a connection in $X$ for amplitude $\delta$ from $\xi_i$ to $\xi_j$. In Section~\ref{sec:model}, theorems~\ref{thm:realisehet} and~\ref{thm:realiseexc}, we present an explicit system whereby any graph $\Gamma$ can be realized as a heteroclinic network or as an excitable network for some small $\delta>0$.

\begin{figure}%
\centerline{\includegraphics[width=15cm]{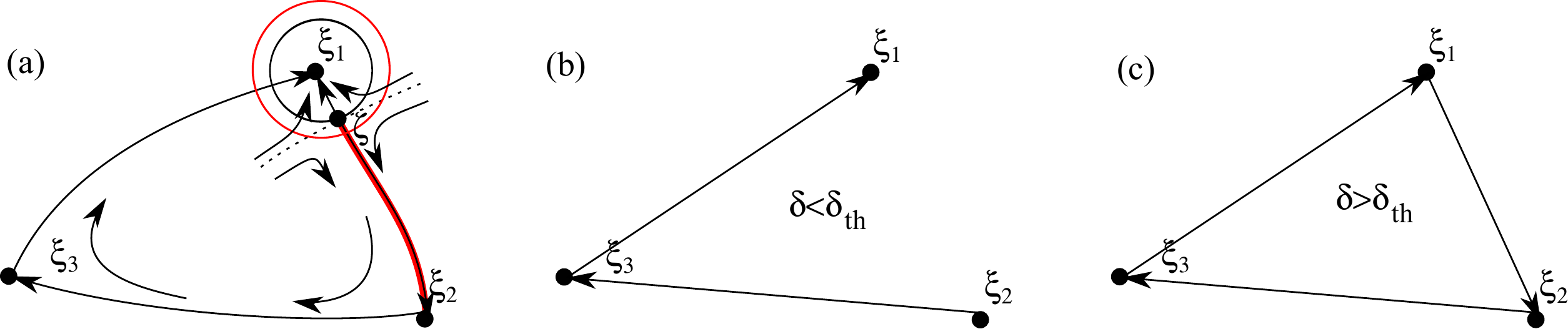}}%
\caption{\red{(a) Schematic diagram showing four equilibria for an example planar vector field. If we examine the three equilibria $\xi_1,\xi_2$ and $\xi_3$ there are heteroclinic connections from $\xi_2$ to $\xi_3$ and from $\xi_3$ to $\xi_1$. There is an excitable connection with threshold $\delta_{th}>0$ that corresponds to the radius of the black circle around $\xi_1$; for any $\delta>\delta_{th}$ (see for example the red circle) there is a connection shown in red from $\xi_1$ to $\xi_2$ with amplitude $\delta$. (b) and (c) show the excitable networks between the $\{\xi_i\}$ for amplitudes $\delta<\delta_{th}$ and $\delta>\delta_{th}$ respectively; note that the heteroclinic network between these equilibria is (b). The threshold corresponds to the distance of the stable manifold (shown by the dotted line in (a)), of the saddle equilibrium $\zeta$, from $\xi_1$.}}%
\label{fig:cartoon2}%
\end{figure}

\red{Note that for a given ODE and set of equilibria there can be a mixture of heteroclinic and excitable connections with different thresholds. Figure~\ref{fig:cartoon2} illustrates this and shows how choosing different amplitudes may give rise to excitable networks with differing topology.}

\red{
There is a subtle difference between existence of a heteroclinic connection between two equilibria and existence of an excitable connection with threshold zero. More precisely one can show the following difference:}

\red{
\begin{lemma}
\label{lem:depth2}
Consider a given ODE and two equilibria $\xi_1$, $\xi_2$.
\begin{itemize}
\item There is a heteroclinic connection from $\xi_1$ to $\xi_2$ if and only if there is a trajectory $x(t)$ such that 
$$
\xi_1=\lim_{t\rightarrow -\infty} x(t),~\mathrm{ and }\ \xi_2=\lim_{t\rightarrow +\infty} x(t).
$$
\item There is an excitable connection from $\xi_1$ to $\xi_2$ with threshold zero if and only if there are trajectories $x_\delta(t)$, $\delta>0$ such that 
$$
\lim_{\delta\rightarrow 0} \liminf_{t\rightarrow -\infty} |x_{\delta}(t)-\xi_1|=0,~\mathrm{ and }\ \xi_2=\lim_{t\rightarrow +\infty} x_\delta(t)~\mathrm{ for\  all}\ \delta>0.
$$
\end{itemize}
\end{lemma}
}

\red{
\proof This follow from considering the definitions of heteroclinic and excitable connection. If there is an excitable connection then for arbitrarily small amplitudes $\delta$ there for each $\delta$ we have a trajectory $x_\delta(t)$ that approaches closely to $\xi_1$ in the limit $t\rightarrow -\infty$; however one may need to choose a different trajectory on reducing $\delta$.
\qed
}

\begin{figure}%
\centerline{\includegraphics[width=45mm]{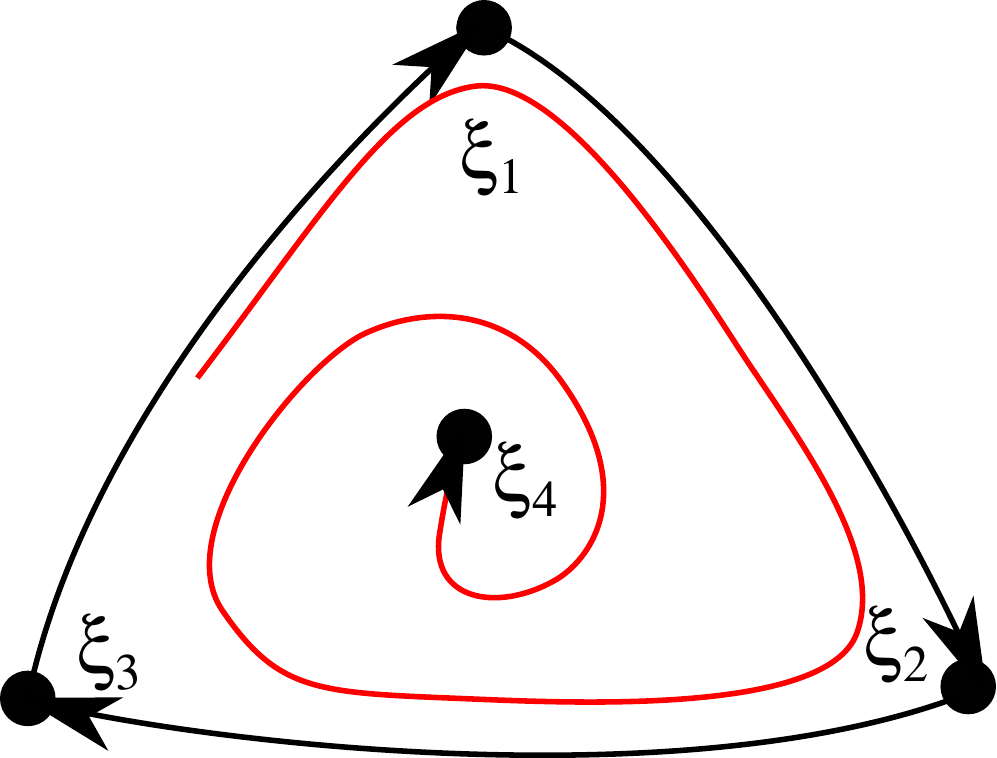}}%
\caption{\red{Schematic diagram illustrating the difference between a heteroclinic connection and an excitable connection with zero threshold. Four equilibria $\xi_i$ are such that there is an excitable (but not a heteroclinic) connection shown in red from any of the $\xi_i$, $i=1,2,3$ to $\xi_4$. Note that the alpha-limit set of the red trajectory contains the heteroclinic cycle between the $\xi_i$, $i=1,2,3$; it is a ``depth two'' connection \cite{ashwin_field_99}.}}%
\label{fig:cartoon1}%
\end{figure}

\red{
Figure~\ref{fig:cartoon1} illustrates that an excitable connection with threshold zero may be a connection of ``depth two'' or greater \cite{ashwin_field_99} even if one can take the same trajectory independent of $\delta$ in Lemma~\ref{lem:depth2}.
}

%
%

\subsection{Example: a cycle of order three}

\label{sec:three_ring_ex}

\begin{figure}%
\centerline{\includegraphics[width=15cm]{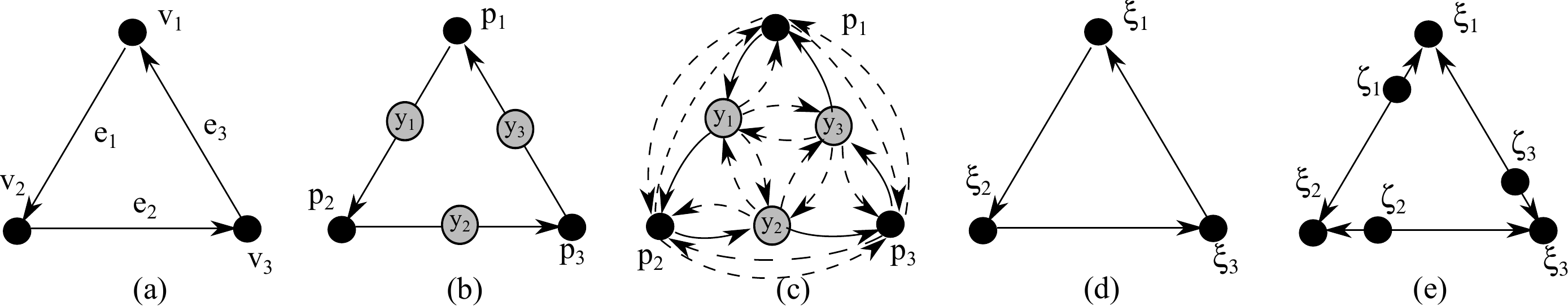}}%
\caption{(a) The cyclic graph: three edges $\e_i$ connect three vertices $\v_i$; (b) schematic coupling architecture of the six-cell network realizing (a). The $p$-cells classify the location when at the vertices of (a) while the $y$-cells only become active during transition between vertices. (c) shows all the connections between the cells; the dashed arrows indicate inhibitory  while the solid arrows indicate excitatory connections. (d,e) schematically show the connections in phase space for this network, where in (d) $\xi_i$ are saddles connected by heteroclinic connections and (e) $\xi_i$ are stable nodes that are connected by excitable connections for amplitude $\delta$, with the separatrices being the stable manifolds of the saddles $\zeta_i$ that are close to the $\xi_i$.}%
\label{fig:C3}%
\end{figure}

We now give a motivating example of how a directed graph can be used to design a coupled cell system. Consider the three-node, three-edge cyclic graph in Figure~\ref{fig:C3}(a). Using two types of dynamical cells we construct a system consisting of six cells (given by equation (\ref{eq:C3system})), as shown in Figure~\ref{fig:C3}(b), where the full coupling between the cells is shown in Figure~\ref{fig:C3}(c). \red{The $p$-cells classify the location (one $p$ cell is active at each vertex in the graph), while the $y$-cells only become active during transition between vertices.} 
\begin{equation}
\begin{split}
\dot{p}_1&=p_1(F(1-p^2)+D(p_1^2p^2-p^4))+E(-y_1^2 p_1p_2 + y_2^2 p_3^2)+\eta_p w_1\\
\dot{p}_2&=p_2(F(1-p^2)+D(p_2^2p^2-p^4))+E(-y_2^2 p_2p_3 + y_1^2 p_1^2)+\eta_p w_2\\
\dot{p}_3&=p_3(F(1-p^2)+D(p_3^2p^2-p^4))+E(-y_3^2 p_3p_1 + y_2^2 p_2^2)+\eta_p w_3\\
\dot{y}_1&=g(y_1,A-B p_1^2+C(y^2-y_1^2))+\eta_y w_4\\
\dot{y}_2&=g(y_2,A-B p_2^2+C(y^2-y_2^2))+\eta_y w_5\\
\dot{y}_3&=g(y_3,A-B p_3^2+C(y^2-y_3^2))+\eta_y w_6
\end{split}
\label{eq:C3system}
\end{equation}
The $w_j$ are white noise processes, $\eta_p$ and $\eta_y$ are noise amplitudes and the function $g$ is given in equation~(\ref{eq:G}). We choose a standard set of parameters (\red{these lie within an open region (\ref{eq:condsABDE}) of suitable parameters} described in section~\ref{sec:model}), and consider the effect of low amplitude noise:
\begin{equation}
A=0.5,~B=1.8,~C=2,~ D=10,~~E=4,~F=2,~\eta_p=\eta_y=10^{-3}.
\label{eq:hetsystemparams}
\end{equation}
The connections between the cells in (\ref{eq:C3system}) are mostly inhibitory (negative feedback), except for the connections shown in Figure~\ref{fig:C3}(c) as solid which represent excitatory connections between selected cells in the sense that they provide positive feedback. Theorem~\ref{thm:realiseexc} from the next section can be used to deduce that there is a heteroclinic cycle as shown schematically in Figure~\ref{fig:C3}(d). Finally, for the same parameters as in~\eqref{eq:hetsystemparams} except choosing $B=1.49$, Theorem~\ref{thm:realisehet} shows that there is an excitable network as shown schematically in  Figure~\ref{fig:C3}(e). 

\begin{figure}%
\begin{center}
\begin{picture}(400,230)
\put(0,0){ \includegraphics[width=15cm]{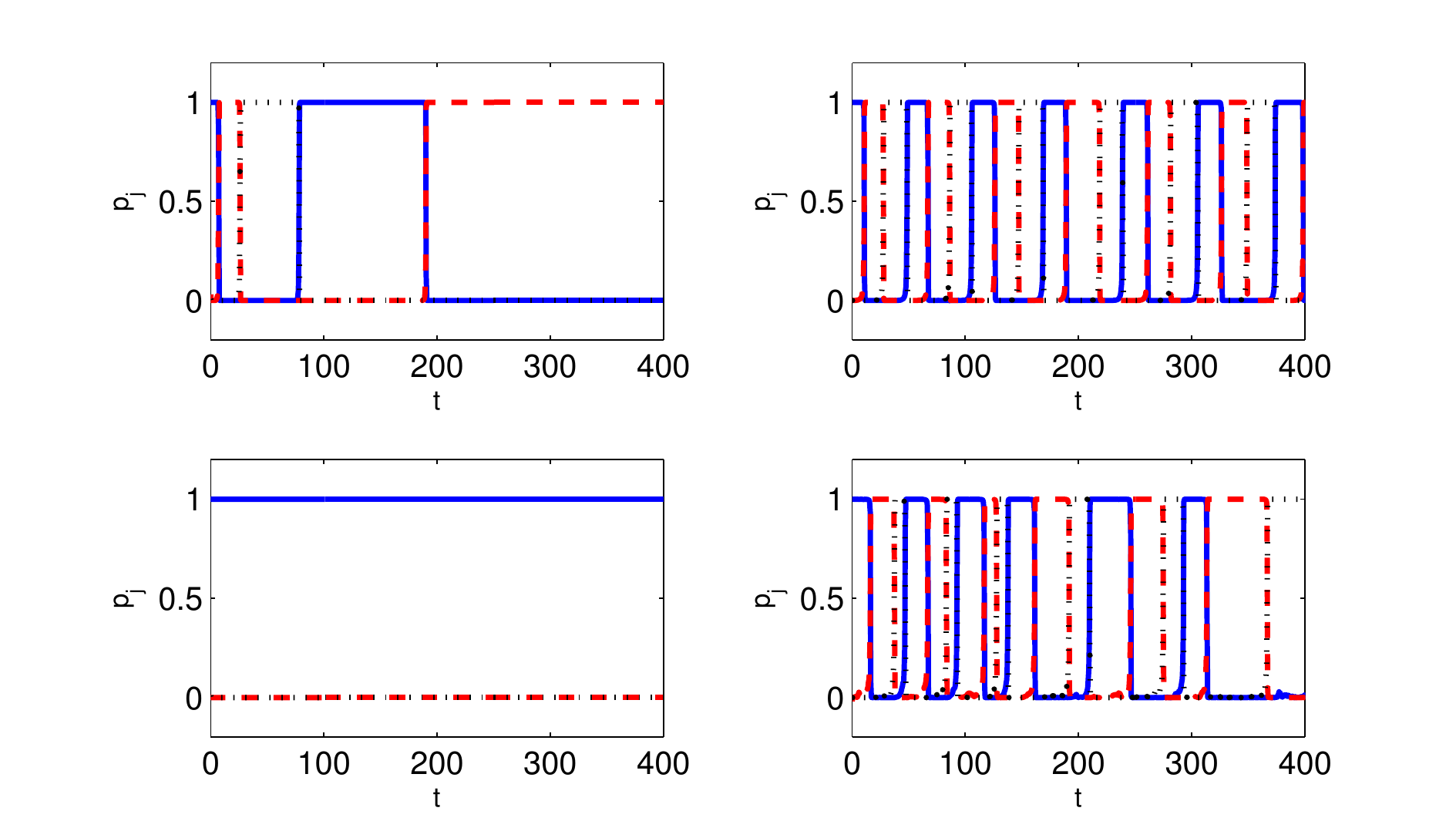}}
\put(25,220){(a)}
\put(25,100){(c)}
\put(225,220){(b)}
\put(225,100){(d)}
 \end{picture}
 \end{center}
\caption{Trajectories for the coupled cell system~\eqref{eq:C3system}, for four different parameter sets. In each panel, $p_1$ is shown by a blue solid line, $p_2$ by a red dashed line, and $p_3$ by a black dotted line. The $y_j$ components are not shown: see figure~\ref{fig:tp1p2}.
 (a) shows the trajectory approaching a heteroclinic network; $B=2.5$ and no noise; $\eta_p=\eta_y=0$. Note the cycling between three states while slowing down typical of a heteroclinic cycle attractor. (b) is as (a) except with non-zero noise; $\eta_p=\eta_y=10^{-3}$. Observe that the slowing down is replaced by an approximate periodicity induced by the noise. In (c), parameters are chosen so there exists an excitable network with no noise; $B=1.49$, $\eta_p=\eta_y=0$. The trajectory approaches a stable equilibrium that depends on initial conditions. (d) is as in (c) except for non-zero noise; $\eta_p=10^{-3}$, $\eta_y=5\times 10^{-2}$. Here, the noise pushes the trajectory over the thresholds at each equilibria and cycling behaviour is seen. {\em (Parameters are as in (\ref{eq:hetsystemparams}), except where stated.)}}%
\label{fig:3cycle}%
\end{figure}

\begin{figure}%
\begin{center}
\begin{picture}(450,200)
\put(0,0){\includegraphics[width=8cm]{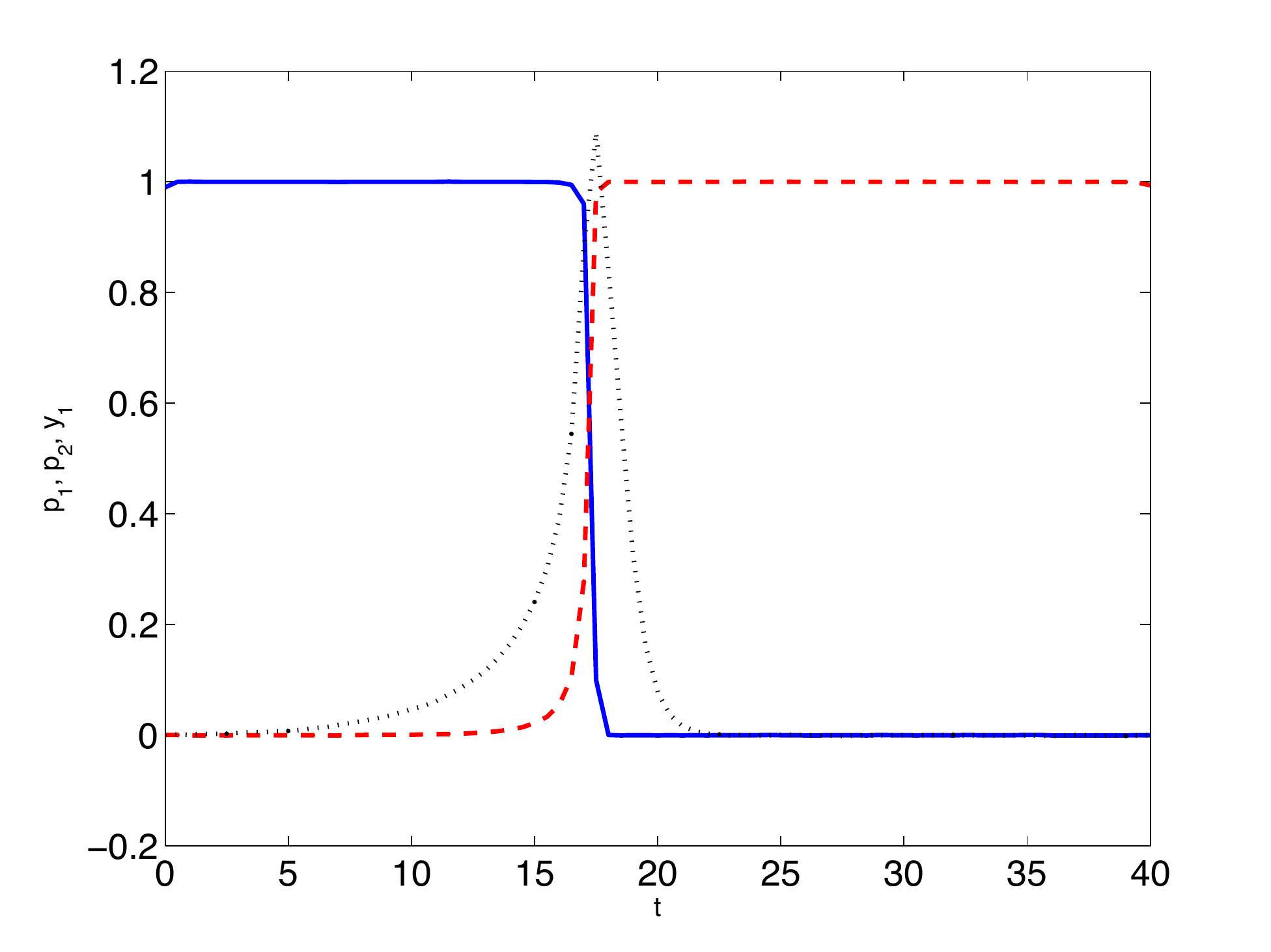}}
\put(225,0){\includegraphics[width=8cm]{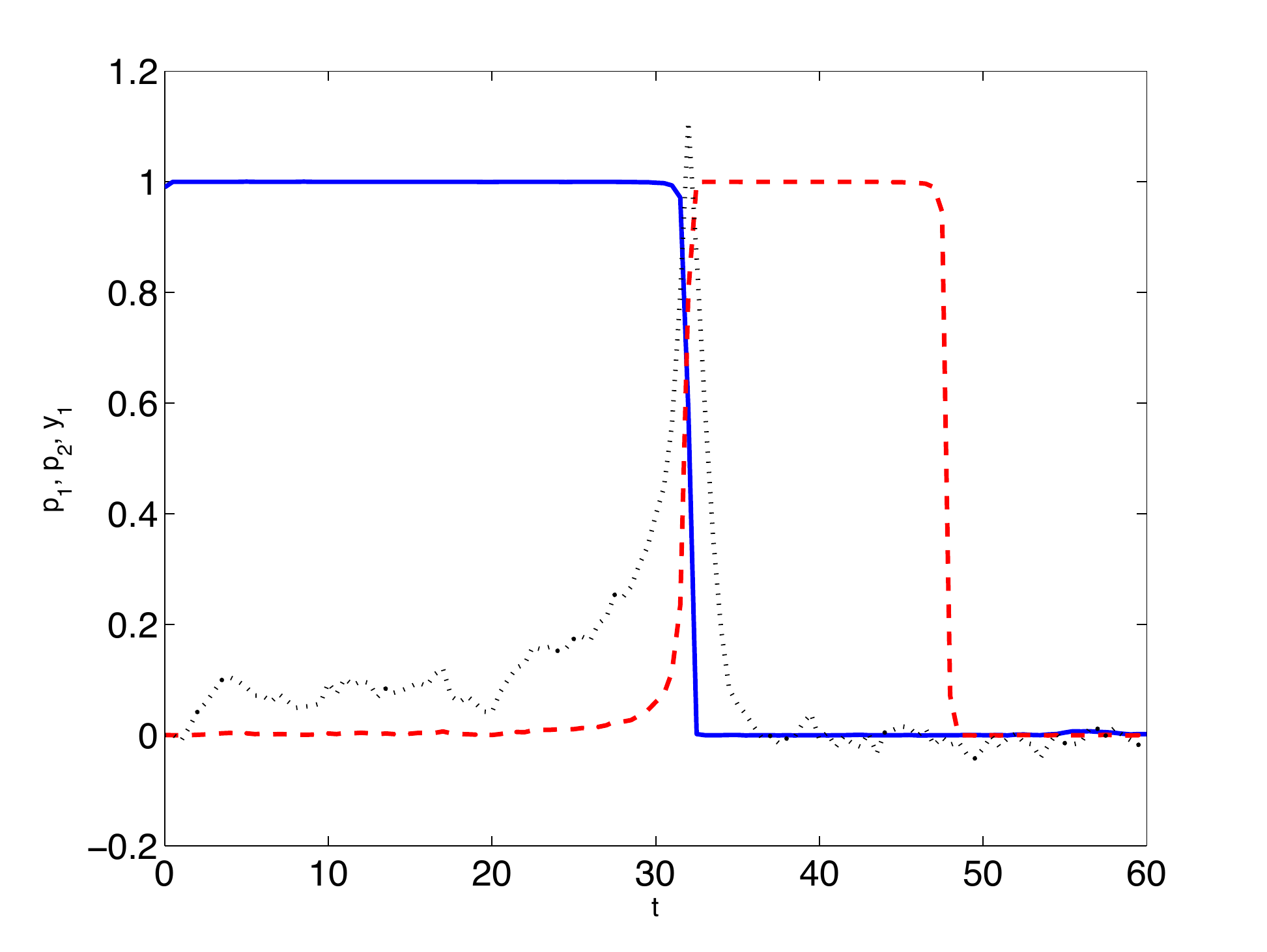}}
\put(0,150){(a)}
\put(220,150){(b)}
\end{picture}
\end{center}
\caption{Time series of transitions between $\xi_1$ and $\xi_2$. In each panel, $p_1$ is shown as a blue solid line, $p_2$ as red dashed and $y_1$ as black dotted. (a) has $B=1.8$, $\eta_p=\eta_y=10^{-3}$, and corresponds to a heteroclinic connection, (b) has $B=1.49$, $\eta_p=10^{-3}$, $\eta_y=3\times 10^{-2}$ and corresponds to an excitable connection.
{\em (Parameters are as in (\ref{eq:hetsystemparams}), except where stated)}}%
\label{fig:tp1p2}%
\end{figure}

Figure~\ref{fig:3cycle} illustrates the attracting behaviour of this system: in the absence of noise (a) and (c), the behaviour of the heteroclinic and excitable networks are quite different. In the presence of noise (b) and (d), they are qualitatively similar due to the trajectories being driven around the network by the noise. In figure~\ref{fig:tp1p2} we show detailed time-series of the system, illustrating the transitions corresponding to edges between the vertices of the directed graph. \red{Note that $y_1$ is switched on during the transition from $\xi_1$ ($p_1=1$) to  $\xi_2$ ($p_2=1$)}.  Throughout this paper, we use a Heun integrator with timestep $h=0.01$ for simulations of the noise-driven systems.

\section{\red{The coupled} system with two cell types}
\label{sec:model}

Consider a directed graph $\Gamma=(\V,\E)$, with vertices and edges defined as in section~\ref{sec:three_ring_ex}. We will realise this both as heteroclinic and as excitable networks in the phase space of a set of coupled cells of two types: the $p$-cells are associated with the vertices $\V$ while the $y$-cells are associated with the edges $\E$. The system we consider has phase space $\mathbb{R}^{n_v+n_e}$, and coordinates $(p,y)=(p_1,\dots,p_{n_v},y_1,\dots,y_{n_e})\in\mathbb{R}^{n_v+n_e}$ governed by:
\begin{equation}
\label{eq:realiseode}
\begin{split}
\frac{d}{dt}p_{j} & =  p_j(F(1- p^2)+ D(p_j^2p^2-p^4))+ E(-Z^{(o)}_j(p,y)+Z^{(i)}_j(p,y))\\
\frac{d}{dt}y_{k} & = g\left(y_{k},A-B p_{\alpha(k)}^2 +C (y^2-y_{k}^2\right))%
\end{split}
\end{equation}
for $j=1,\cdots,n_v$ and $k=1,\cdots,n_e$, where $p^2=\sum_{j=1}^{n_v} p_j^2$, $p^4=\sum_{j=1}^{n_v} p_j^4$, $y^2=\sum_{j=1}^{n_e} y_j^2$ and $A,B,C,D,E,F$ are constants. The function $g$ is defined by
\begin{equation}
g(y_{k},\lambda)= -y_{k}\left((y_{k}^2-1)^2+\lambda\right)
\label{eq:G}
\end{equation}
while the inputs to the $p_j$ cells from the $y$ cells are:
\begin{equation}
\begin{split}
Z^{(o)}_j(p,y) &= \sum_{\{k~:~\alpha(k)=j\}} -y_k^2p_{\omega(k)}p_j\\
Z^{(i)}_j(p,y) &= \sum_{\{k'~:~\omega(k')=j\}} y_{k'}^2p_{\alpha(k')}^2.
\end{split}
\end{equation}
Equations~\eqref{eq:realiseode} have equilibria at $\xi_j=(0,\dots,1,\dots,0)\in\mathbb{R}^{n_v+n_e}$ for $j=1,\dots,n_v$ where the ``$1$'' is in the $j$th position. That is, the equilibria are at points corresponding to unit vectors where one of the $p_j$ is non-zero.

\begin{figure}%
\centerline{\includegraphics[width=6cm]{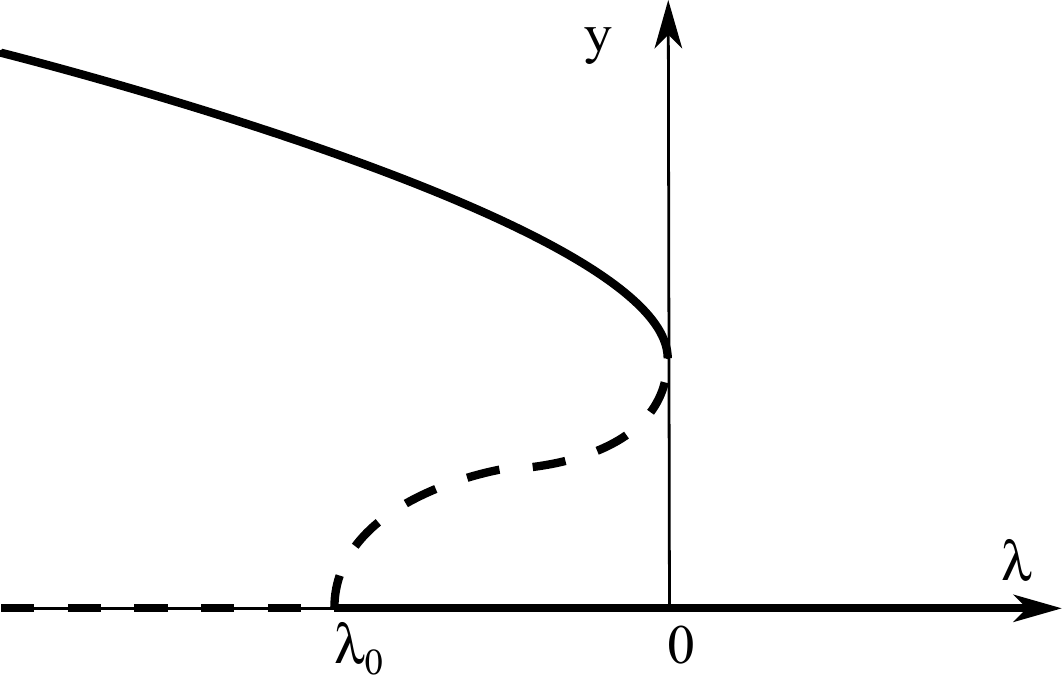}}%
\caption{Bifurcation diagram of $\dot{y}=g(y,\lambda)$ (given in (\ref{eq:G})) for $y\geq 0$ and $\lambda\in\R$. Note that there is a region of bistability between the pitchfork bifurcation at $\lambda=\lambda_0:=-1$, $y=0$ and the saddle-node bifurcation at $\lambda=0$, $y=1$.}%
\label{fig:Gbifs}%
\end{figure}

Note that $\dot{y}=g(y,\lambda)$ has a hysteresis loop that can be switched by changing $\lambda$ through the interval $[\lambda_0,0]$ where $\lambda_0:=-1$; see Figure~\ref{fig:Gbifs}; in this sense, perturbations that reduce $\lambda$ are excitatory while those that increase $\lambda$ are inhibitory. The coupling and the choice of parameters will be made so as to construct a network in phase space where each connection goes once around a hysteresis loop within a subspace $P_{\ell}$ (defined in the following section). 

\subsection{Dynamics of the model}

System (\ref{eq:realiseode}) has symmetries $\Z_2^{(k)}$ given by $y_k\mapsto -y_k$ for each $k$ and so the system is equivariant under the action of the group
$$
\Sigma=\prod_{k=1}^{n_e} \Z_2^{(k)}.
$$
We prove the existence of networks in phase space that realise the given graph and are robust to perturbations that respect this symmetry. To this end we denote by $\Sigma_{\ell}$ the subgroup of $\Sigma$ corresponding to $\Z_2^{(\ell)}$, and define the following subspaces of phase space
$$
\begin{array}{c}
Y_{\ell}:=\mbox{fix}(\Sigma_{\ell})=\{(p,y)~:~ y_{\ell}=0\}\\
W_{\ell}:=\bigcap_{k\neq \ell} \mbox{fix}(\Sigma_{k})=\{(p,y)~:~y_{k}=0~\mbox{ if }k\neq \ell\}
\end{array}
$$
and
$$
P_{\ell}:=\{(p,y)~:~y_k=0~\mbox{ if }k\neq \ell~\mbox{ and }p_j=0~\mbox{ if }j\neq\alpha(\ell)\mbox{ or }\omega(\ell)\}\\
$$
for $\ell=1,\ldots,n_e$. The sets $Y_{\ell}$ and $W_{\ell}$ are invariant for all $\Sigma$-equivariant perturbations, while the $P_{\ell}\subset W_{\ell}$ 
are invariant for the flow generated by system (\ref{eq:realiseode}) but not for arbitrary $\Sigma$-equivariant perturbations. However, we will show that there are connecting orbits in the $P_{\ell}$ subspaces that are robust to small $\Sigma$-equivariant perturbations that preserve the invariance of the larger subspace $W_{\ell}$. We define 
\begin{equation}
\cS:=\{ (p,y)~:~|p|^2=1\}\equiv S^{n_v\red{-1}}\times \R^{n_e} \label{eq:S}
\end{equation}
which is an $n_v\red{-1}$-dimensional sphere in the $p$-coordinates. This is invariant and normally attracting for $F>0$, and so persists for appropriate choice of the parameters. We interpret the parameters $A,B,C,D,E,F$ in~\eqref{eq:realiseode} as follows:
\begin{itemize}
\item The constant $A$ determines the default dynamics of the variables $y_k$: we assume $A\geq 0$ so that the equilibria $\xi_j$ are globally stable for $B=C=0$.
\item The constant $B>0$ determines how much $\xi_j$ is destabilised by there being a connection from that state. Let $\alpha(k)=j$. Then if $B>A+1$,  $\xi_j$ will be linearly unstable in the $y_k$ direction. If $A+1>B>0$ then the state $\xi_j$ will be linearly stable but excitable in the $y_k$ direction.
\item The constant $C>0$ determines the mutual inhibition of the $y_k$ variables and suppresses more than one hysteresis loop becoming active at any time. 
\item The constant $D>0$ sets the rate of attraction to the equilibria $\xi_j$ in directions tangent to $\cS$.
\item The constant $E>0$ is set to lie within a range (relative to $D$) so that when one of the $y_k$ is active then there is a connection from $\xi_{\alpha(k)}$ to $\xi_{\omega(k)}$ (see Figure~\ref{fig:regions}, and equation~\eqref{eq:condsABDE} for details).
\item The constant $F>0$ sets the rate of attraction of the $p$ dynamics towards $\cS$.
\end{itemize}

\begin{lemma}\label{lem:sphere}
\red{The system (\ref{eq:realiseode}) has an invariant set $\cS$ (defined in~\eqref{eq:S}); for $F>0$ this set attracts a neighbourhood of $\cS$.}
\end{lemma}

\proof
We show that if $\Lambda:=p^2=\sum_{j=1}^{n_v} p_j^2$ then $\Lambda\rightarrow 1$ as $t\rightarrow \infty$ for typical initial conditions. Note that
\begin{eqnarray*}
\frac{1}{2} \frac{d}{dt} \Lambda &=& \sum_{j} p_j\dot{p}_j \\
&=& F\sum_{j=1}^{n_v} p_j^2-Fp^2\sum_{j=1}^{n_v} p_j^2 + D\left(p^2 \sum_{j=1}^{n_v} p_j^4 - p^4\sum_{j=1}^{n_v} p_j^2\right)\\
&& + E\left( \sum_{k=1}^{n_e} -y_k^2 p_{\alpha(k)}^2p_{\omega(k)}+\sum_{k'=1}^{n_e} y_{k'}^2p_{\alpha(k')}^2p_{\omega(k')}\right)\\
& = & F\Lambda(1-\Lambda).
\end{eqnarray*}
Hence on a timescale determined by $F>0$, we typically have $\Lambda\rightarrow 1$ as $t\rightarrow \infty$. The only initial conditions where this is not the case will have $p_j=0$ for all $j$. 
\qed

\subsection{Realisation of a graph as a heteroclinic network}

The following lemma shows that for an open region in parameter space the dynamics of system~\eqref{eq:realiseode} embeds the graph $\Gamma$ as a heteroclinic network.

\begin{lemma}
\label{lem:connects}
Consider the system (\ref{eq:realiseode}) with equilibria at $\xi_j$ for $j=1,\ldots,n_v$. There is an open set of $A,B,C,D,E,F$  such that for each $\ell=1,\ldots,n_e$ there is a connecting orbit from $\xi_{\alpha(\ell)}$ to $\xi_{\omega(\ell)}$ within the three dimensional invariant subspace $P_\ell$.
\end{lemma}

\proof
For ease of exposition, and without loss of generality, suppose $\ell=1$, $\alpha(\ell)=1$ and $\omega(\ell)=2$. The system within the invariant subspace $P_{1}$ can be written
\begin{eqnarray}
\dot{p}_1&=&p_1(F(1-p^2)+D(p_1^2p^2-p^4))-Ey_1^2 p_1 p_2\nonumber\\
\dot{p}_2&=&p_2(F(1-p^2)+D(p_2^2p^2-p^4))+Ey_1^2 p_1^2\label{eq:P1eq}\\
\dot{y}_1&=&g(y_1,A-B p_1^2)\nonumber
\end{eqnarray}
where $p^2=p_1^2+p_2^2$. Then the linearized stability within $P_{1}$ is given by
{\small $$
\left(\begin{array}{ccc} 
F(1-3p_1^2-p_2^2) + D(3p_1^2p^2-p_2^4) - E p_2 y_1^2
	& p_1(D(2p_1^2p_2-4p_2^3) -2Fp_2 - E y_1^2)
	& - 2E p_1 p_2 y_1 \\
p_1(D(2p_2^3-4p_2p_1^2)-2Fp_2+ 2E y_1^2)
	& F(1-3p_2^2-p_1^2) + D(3p_1^2p^2-p_1^4)
	& 2E p_1^2 y_1\\
-y_1 ( A-2 Bp_1) 
	& 0 
	& g'(y_1,A-Bp_1^2)
\end{array}\right)
$$}
where $g'(y,\lambda):=\dfrac{dg}{dy}(y,\lambda)$. For the point $\xi_1=(1,0,0)$ this becomes
$$
\left(\begin{array}{ccc} 
-2F &  0 &  0\\
 0	& -D &  0\\
 0  &  0 &  B-1-A
\end{array}\right)
$$
while for $\xi_2=(0,1,0)$ it becomes
$$
\left(\begin{array}{ccc} 
 -D	&   0 & 0 \\
  0 & -2F & 0 \\
  0 &  0 	& -1-A
\end{array}\right).
$$
Hence, we choose $F>0$, $D>0$, $A>0$ and $B>1+A$ so that $\xi_1$ is a saddle with unstable direction $(0,0,1)$ and $\xi_2$ is a stable node. Observe that for this choice both $\xi_1$ and $\xi_2$ are hyperbolic and that all other eigenvalues in the direction of other $p_k$ are $-D$ and therefore stable.

As the subset $\mathcal{C}=\cS\cap P_{1}$ (where $p_1^2+p_2^2=1$) is attracting and invariant (by Lemma~\ref{lem:sphere}), we consider the dynamics on $\mathcal{C}$ parametrized by $(\theta,y_1)$ where $p_1=\cos \theta$, $p_2=\sin \theta$ and $\theta\in[0,2\pi)$. From equation (\ref{eq:P1eq}) we have
\begin{eqnarray*}
\frac{d\theta}{dt} &=&  p_1p_2 D(p_2^2-p_1^2) + E p_1(p_1^2+p_2^2) = D \sin \theta\cos \theta (\sin^2(\theta)-\cos^2(\theta)) + E y_1^2 \cos \theta\\
& = & -\frac{D}{4} \sin 4\theta + E y_1^2 \cos \theta
\end{eqnarray*}
so that in this subspace we have
\begin{eqnarray}
\frac{d\theta}{dt} &=& -\frac{D}{4} \sin 4\theta + E y^2 \cos \theta \label{eq:twod1}\\
\frac{dy}{dt} &=& - y ((y^2-1)^2+A-B\cos^2\theta) \label{eq:twod2}
\end{eqnarray}
where we drop the subscript from $y_1=y$ for notational convenience.

\begin{figure}%
\centerline{\includegraphics[width=15cm]{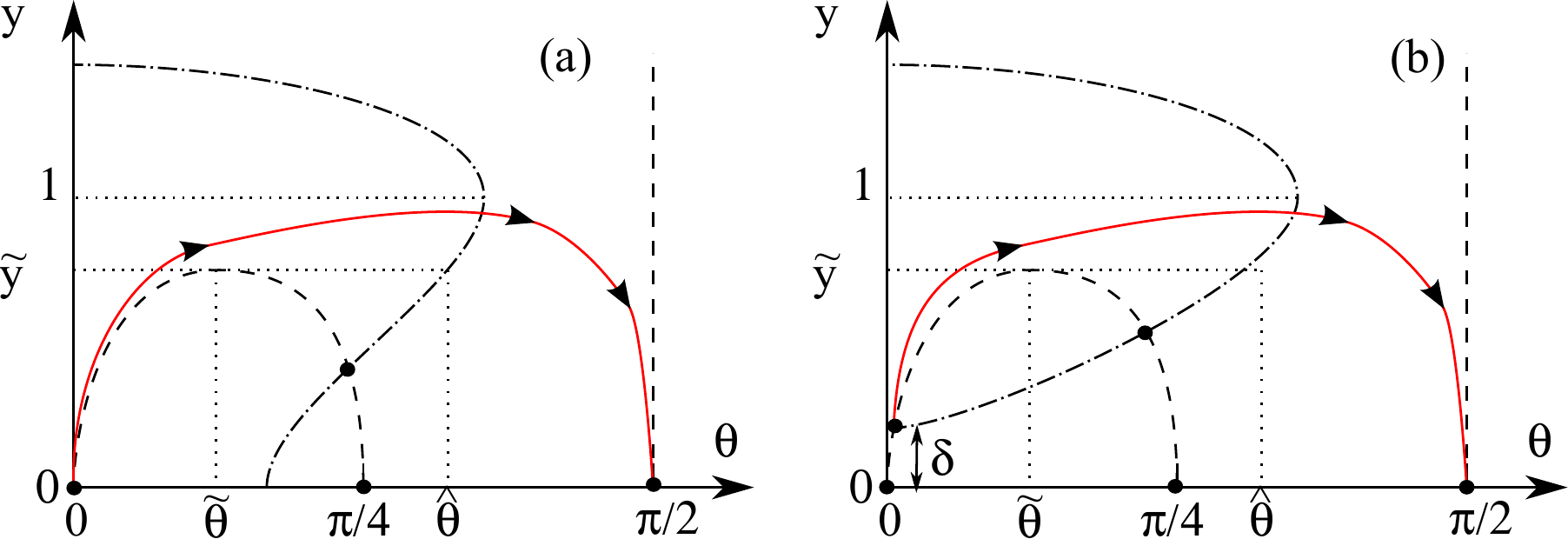}}%
\caption{ The phase plane for $(\theta,y)$ showing the nullclines in the region $\theta\in[0,\pi/2]$. (a) The dashed line shows $\dot{\theta}=0$ while the dash-dotted line shows $\dot{y}_1=0$; equilibria are indicated with disks. If the nullclines have the given topology and are such that $\tilde{y}<1$ and $\pi/4<\hat{\theta}<\pi/2$ then there will be a heteroclinic connection as shown (schematically as a solid red line). This can be achieved by choosing constants that satisfy (\ref{eq:condsABDE}); see text for more details. (b) For larger values of $B$ the heteroclinic connection becomes an excitable connection (schematically as a solid red line) for amplitude $\delta$ (schematically as a solid red line) and a new saddle equilibrium with $y\neq 0$ appears.}
\label{fig:connection}%
\end{figure}

We use~\eqref{eq:twod1} and~\eqref{eq:twod2} to deduce conditions on the parameters $A,B,D,E$ that guarantee existence of a saddle-to-sink connection from $(\theta,y)=(0,0)$ to $(\theta,y)=(\pi/2,0)$ corresponding to existence of a heteroclinic connection from $\xi_1$ to $\xi_2$ within $P_1$. Note that the $\dot{\theta}=0$ nullclines are at $\cos \theta=0$ or at $y^2=D \sin \theta(2\cos^2\theta-1)$ (shown by the dashed lines in Figure~\ref{fig:connection}).  For $D>0$ the latter has a unique maximum in $[0,\pi/2]$ at 
$$
\tilde{\theta} := \sin^{-1}\left(\frac{1}{\sqrt{6}}\right),~~\tilde{y} :=\sqrt{\frac{D\sqrt{6}}{9E}}.
$$
If $D>0$ and $E>0$, $\dot{\theta}>0$ whenever $y>\tilde{y}$.
We require that $\tilde{y}<1$ in order to rule out the possibility of any equilibria in $y>1$.

\begin{figure}%
\centerline{\includegraphics[width=8cm]{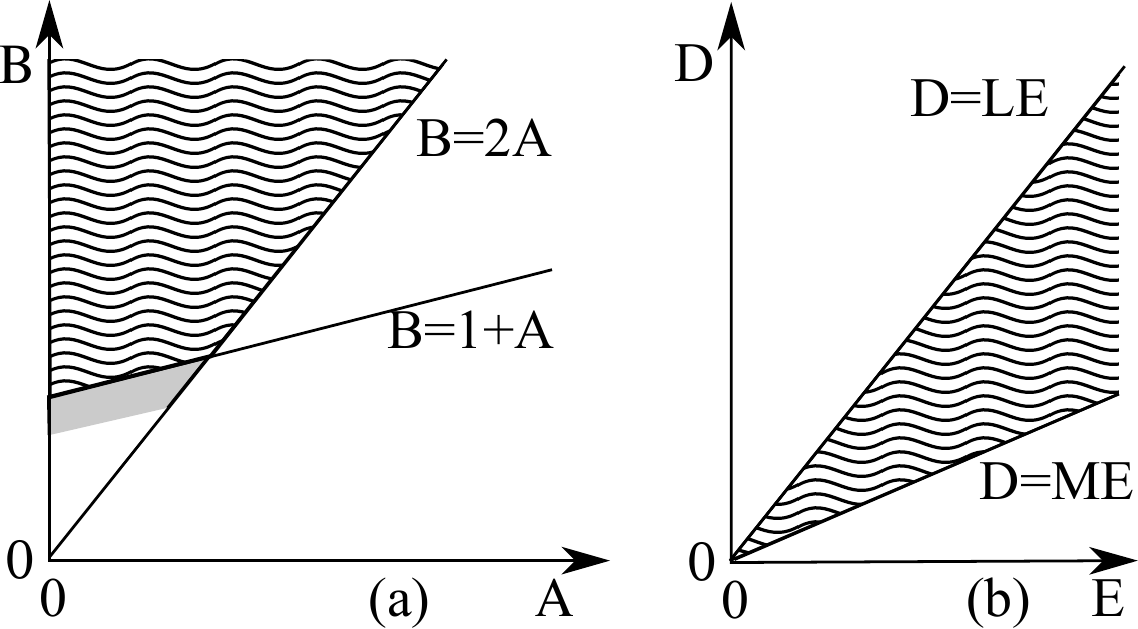}}%
\caption{The parameter region that satisfies the conditions (\ref{eq:condsABDE}) that permit a heteroclinic realization are illustrated here: (a) $(A,B)$ can be chosen from the region shown in wavy lines. (b) For each choice there is an $M<L$ such that $D,E$ can be chosen from the region shown in wavy lines, where $L=9/\sqrt{6}$ and $M= L(1-\sqrt{B/2-A})$ which depends on $A$ and $B$. The grey shaded region in (a) can be added to the allowable conditions if we permit excitable realizations with small $\delta>0$. Note that not all of the networks are attracting, but for small enough $B$ and suitable choices of $C,F$, numerical simulations indicate that they are.}
\label{fig:regions}%
\end{figure}

The $\dot{y}=0$ nullclines are at $y=0$ and $B\cos^2\theta=(1-y^2)^2+A$. If $B>1+A$ the latter curve has a minimum in $\theta$ at $y=0$ and maxima in $\theta$ at $y=\pm 1$ (dashed-dotted line in figure~\ref{fig:connection}(a)). Suppose that the line $y=\tilde{y}$ hits the $y$-nullcline first in $[0,\pi/2]$ at $\hat{\theta}$; this is given by
$$
\cos^2 \hat{\theta} = \frac{1}{B}\left( \left(1-\frac{D\sqrt{6}}{9E}\right)^2 + A \right).
$$
If $\pi/4< \hat{\theta}<\pi/2$ and $0<\tilde{y}<1$ then Figure~\ref{fig:connection}(a) shows that there will be a connection as desired; hence some sufficient (but by no means necessary) conditions for there to be a connection can be expressed as:
\begin{equation}
\label{eq:condsABDE}
0<A<\frac{B}{2},~~1+A<B,~~0<D, ~~0<E~~\mbox{ and }~~
\frac{9E}{\sqrt{6}}\left(1-\sqrt{\frac{B}{2}-A}\right)<D<\frac{9E}{\sqrt{6}}
\end{equation}
Note that $C,F$ do not affect this argument; however $C$ needs to be chosen to positive and large enough to avoid spurious connections to other stable dynamics and $F$ needs to be positive for Lemma~\ref{lem:sphere} to hold.  Figure~\ref{fig:regions} illustrates that this set is non-empty and open.
\qed

To illustrate the effect of one component of the $y$ dynamics becoming active and giving a connection, we show in Figure~\ref{fig:pdyns} the dynamics in $p_1$ and $p_2$ for (a) $y_1=0$ and (b) $y_1\approx 1$.  We summarise the construction above in the following Theorem:

\begin{figure}%
\centerline{\includegraphics[width=10cm]{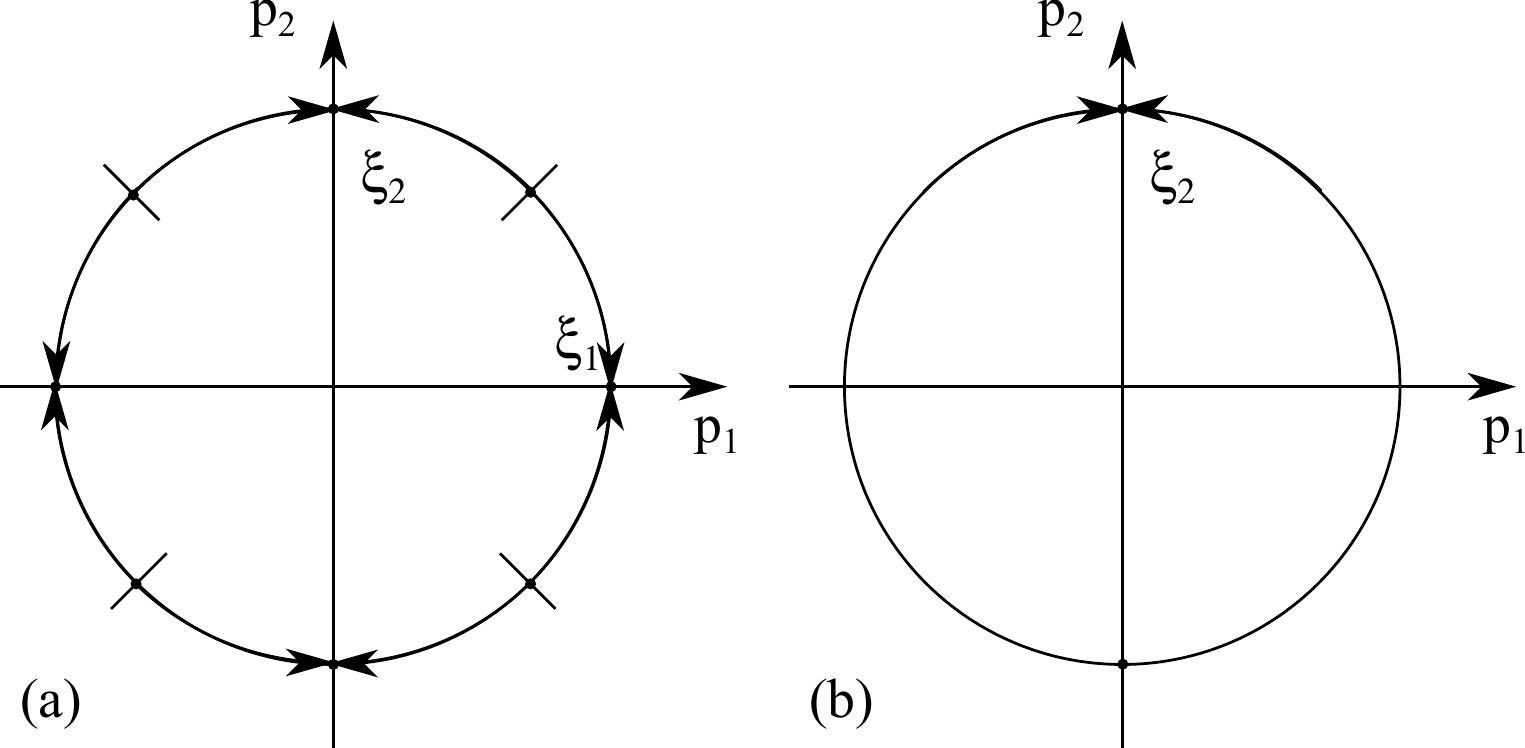}}%
\caption{The dynamics of $p_1$ and $p_2$ on the unit circle in the case that $y_1$ is associated with a connection from $\v_1$ to $\v_2$, i.e. in the case $\alpha(1)=1$, $\omega(1)=2$. (a) For $y=0$ note that both $\xi_1$ and $\xi_2$ are stable nodes; note that $\xi_1$ is unstable in the $y_1$ direction. (b) For $y_1\approx 1$. As $y_1$ is increased from $0$, the terms multiplied by $E$ in (\ref{eq:realiseode}) removes $\xi_1$ (and all symmetric images) in a series of bifurcations.}
\label{fig:pdyns}%
\end{figure}

\begin{theorem}
\label{thm:realisehet}
Given any finite directed graph, there is a non-empty and open set of parameter values $A,B,C,D,E,F$ such that the system (\ref{eq:realiseode}) realises this graph as a heteroclinic network in a way that is robust to all perturbations to the equations that preserved the symmetries $\Sigma$.
\end{theorem}

\proof
The previous calculations and Lemma~\ref{lem:connects} show that for the parameter region identified in (\ref{eq:condsABDE}) and with $C>0$, $F>0$ there are hyperbolic saddles $\{\xi_i\}$ that are connected by heteroclinic connections in the three-dimensional subspaces $P_{\ell}$ for each $\ell$. These connections are robust to perturbations that preserve the symmetry $\Sigma$ because within the fixed point subspace  $Y_{\ell}$ the equilibrium $\xi_{\alpha(\ell)}$ has one unstable direction and all other directions are stable while $\xi_{\omega(\ell)}$ is a sink. 
Moreover $P_{\ell}\subset Y_{\ell}$ and so there is a connection that is of saddle-sink type. Transversality of this connection means that it is robust to $\Sigma$-equivariant perturbations.
\qed

\red{We conjecture that, in the case of a strongly connected graph $\Gamma$, for some open subset of the set of parameters in Theorem~\ref{thm:realisehet} this heteroclinic network realises the graph as part of an asymptotically stable attractor that is a compact, chain recurrent invariant set. As in \cite{AshPos13}, the large ``embedding attractor'' will typically contain extra equilibria and connections but we conjecture that the proportion of time that typical trajectories visit equilibria that do not correspond to those in $\Gamma$ will be very small and may go to zero as noise amplitude decreases; the larger attractor may be ``invisible'' \cite{Ilyashenko2010} except on a subset that corresponds to an embedding of $\Gamma$.} 

Some numerical evidence for these conjectures is given in the next section for a specific example.  In brief justification, if $\xi_1$ has connections to $\xi_2$ and $\xi_3$ via $y_1$ and $y_2$ then for large enough $C$, almost every trajectory on $W^u(\xi_1)$ is a connection to one of $\xi_{2}$ or $\xi_3$. This is suggested by the dynamics of $y_1$, $y_2$ which are governed by:
\begin{eqnarray}
\frac{dy_1}{dt} &=& - y_1 ((y_1^2-1)^2+1+A-Bp_1^2+C y_2^2)\\
\frac{dy_2}{dt} &=& - y_1 ((y_1^2-1)^2+1+A-Bp_1^2+C y_1^2).
\end{eqnarray}
Fixing $p_1=1$ and $1+A-B=0$ and examining the phase plane for this system, if $C>2/3$ then all orbits are bounded in forwards time and the only attractors for this system are in $y_1=0$ and $y_2=0$. This is preserved for $1+A-B$ close to zero.

\subsection{Realisation of a graph as an excitable network}

We give an additional result to show that a given graph can be realized as an excitable network for amplitude $\delta>0$ for a range of parameter values:

\begin{theorem}
\label{thm:realiseexc}
Given any finite directed graph, there is an open set of parameter values $A,B,C,D,E,F$ determining a minimum amplitude (threshold) $\delta_{th}>0$ such that (\ref{eq:realiseode}) realises this graph as an excitable network with amplitude \red{$\delta$} for $\delta>\delta_{th}$ but not for $\delta<\delta_{th}$. 
\end{theorem}

\proof
We choose $A,B,C,D,E,F$ as in Theorem~\ref{thm:realisehet} except we do not require $1+A-B<0$ meaning that the $y_1$ nullcline may be detached from $y_1=0$; see Figure~\ref{fig:connection}(b). That is, the curve $B\cos^2\theta=(1-y^2)^2+A$ is undefined for $0<y<\hat{\delta}$, where it is simple to show that 
\[
\hat\delta=\sqrt{1-\sqrt{B-A}};
\]
see the dot-dashed curve in Figure~\ref{fig:connection}(b). In this case $\xi_1$ is a sink within the invariant subspace $P_1$ and there is a nearby saddle $\zeta_1$ whose stable manifold forms part of the boundary of $W^s(\xi_1)$. Let $\delta_0$ be the smallest distance from $\xi_1$ to the stable manifold of $\zeta_1$.

More precisely, if we consider $1+A-B=\nu>0$, $\nu \ll 1$,  then we can estimate $\delta_{th}$, the closest approach of $W^s(\zeta_1)$ to $\xi_1$ by $\hat{\delta}$, the point where the $y_1$ nullcline intersects $\theta=0$. This gives
\begin{equation}
\label{eq:delta}
\delta_{th}= \sqrt{\frac{\nu}{2}} + O(\nu).
\end{equation}
for small $\nu$; in other words, for $\delta>\delta_{th}$ there will be a connection for amplitude $\delta$ while for $\delta<\delta_{th}$ we have $B_{\delta}(\xi_1)\subset W^s(\xi_1)$ and there is no connection for amplitude $\delta$.
\qed

In the case of multiple outgoing directions there will be multiple directions with a threshold of $\delta_{th}$ and a similar argument to that following Theorem~\ref{thm:realisehet} suggests that for large enough $C>0$ and $\delta>\delta_{th}$, almost all points in $B_{\delta}(\xi_k)$ are either in $W^s(\xi_k)$ or in $W^s(\xi_j)$ for some $\xi_j$ that is connected via one of these outgoing directions.

\section{Design of a system possessing a Kirk--Silber cycle}
\label{sec:examples}

One of the simplest examples of a network that shows competition between two heteroclinic cycles is the network of Kirk and Silber \cite{KS94} where two order-three cycles (similar to that in Figure~\ref{fig:3cycle}) share a common edge. This is a useful system to understand how the system switches at the ``decision point'' in response e.g. to noise of differing amplitudes in different components. In a forthcoming paper \cite{AshPos15} we explore the statistics of the switching process in terms of escape processes simultaneously along a number of heteroclinic or excitable connections; here we indicate some of the issues in this example.

\begin{figure}%
\centerline{\includegraphics[width=10cm]{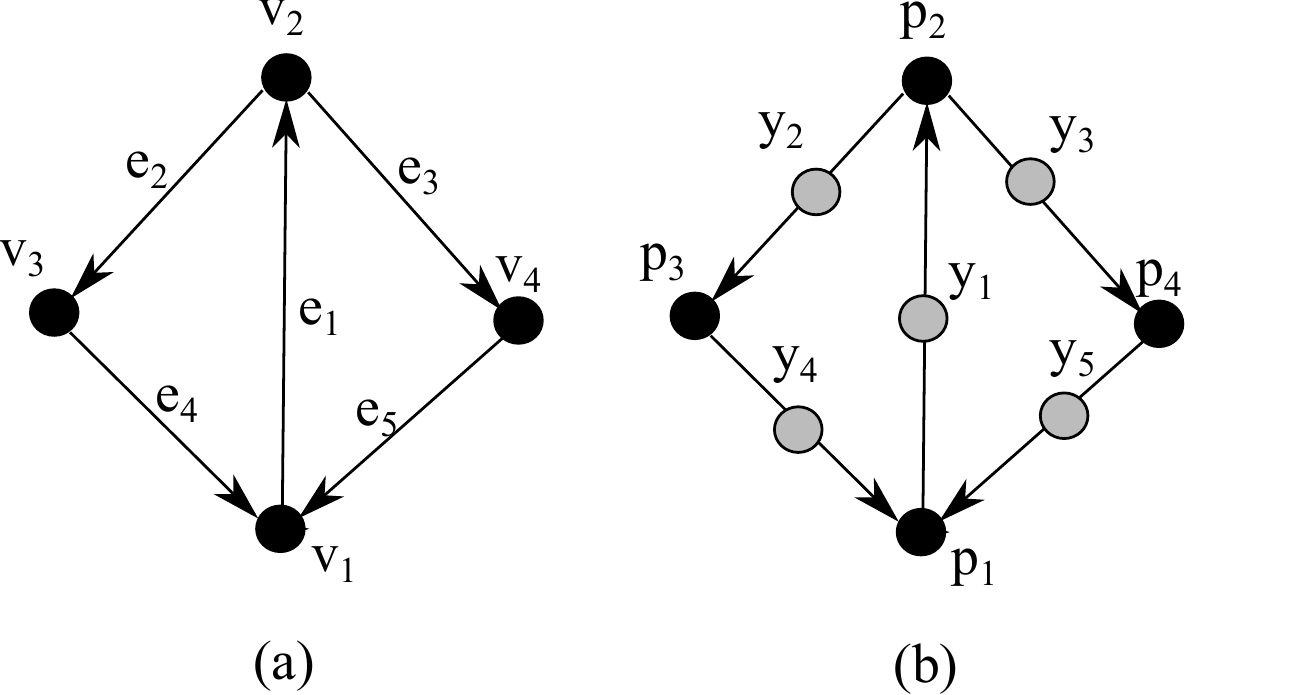}}%
\caption{(a) The Kirk--Silber network: five edges $\e_i$ connect four vertices $\v_i$; (b) schematic of the coupling architecture of the nine-cell network realizing (a); the globally inhibitory connections are not shown.
}%
\label{fig:KS}%
\end{figure}

\begin{figure}%
\begin{center}
\begin{picture}(400,400)
\put(0,0){ \includegraphics[width=16cm]{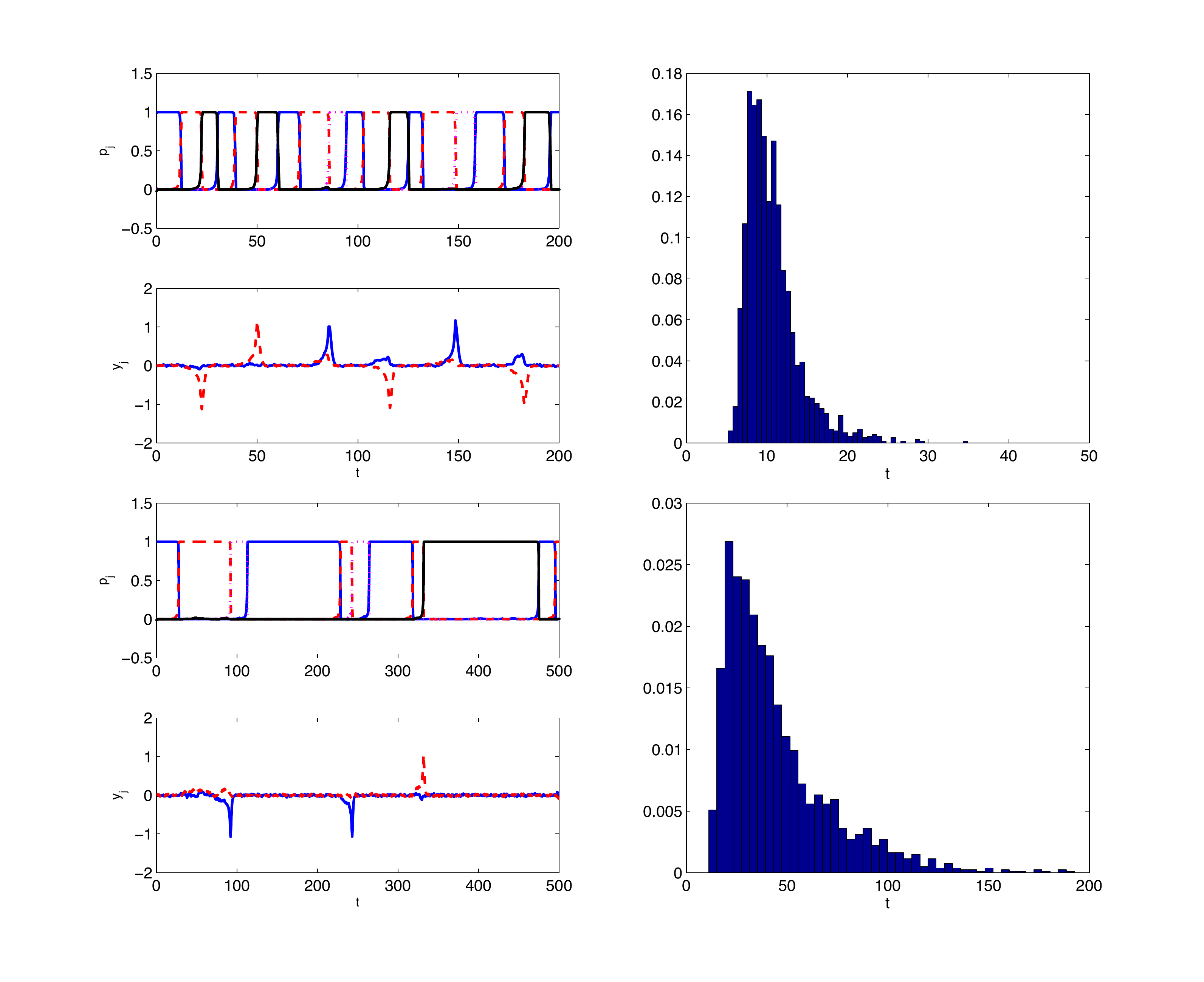}}
\put(12,330){(a)}
\put(12,250){(b)}
\put(12,170){(c)}
\put(12,90){(d)}
\put(230,330){(e)}
\put(230,170){(f)}
 \end{picture}
 \end{center}
\caption{Time series and histograms of residence times for the Kirk--Silber example in~\eqref{eq:kssystem}. (a), (b) and (e) are for the heteroclinic case (parameters as in  (\ref{eq:hetsystemparams}), but with $\eta_j=3\times10^{-5}$ ($j=1,\dots,5$)). (c), (d) and (f) are for the excitable case (parameters as in  (\ref{eq:hetsystemparams}), but with $B=1.49$ and $\eta_j=3\times10^{-5}$ ($j=1,\dots,5$)). (a) and (c) show time series for the $p_j$ (linestyles are: 
$p_1$: blue solid, $p_2$: red dashed, $p_3$ magenta dotted, $p_4$: black solid); (b) and (d) show time series for the $y_j$, we only show $y_ 2$ (blue solid line) and $y_3$ (red dashed line). (e) and (f) show histograms of residence times near equilibria for each case for a much longer time series. 
}%
\label{fig:KS-TS}%
\end{figure}
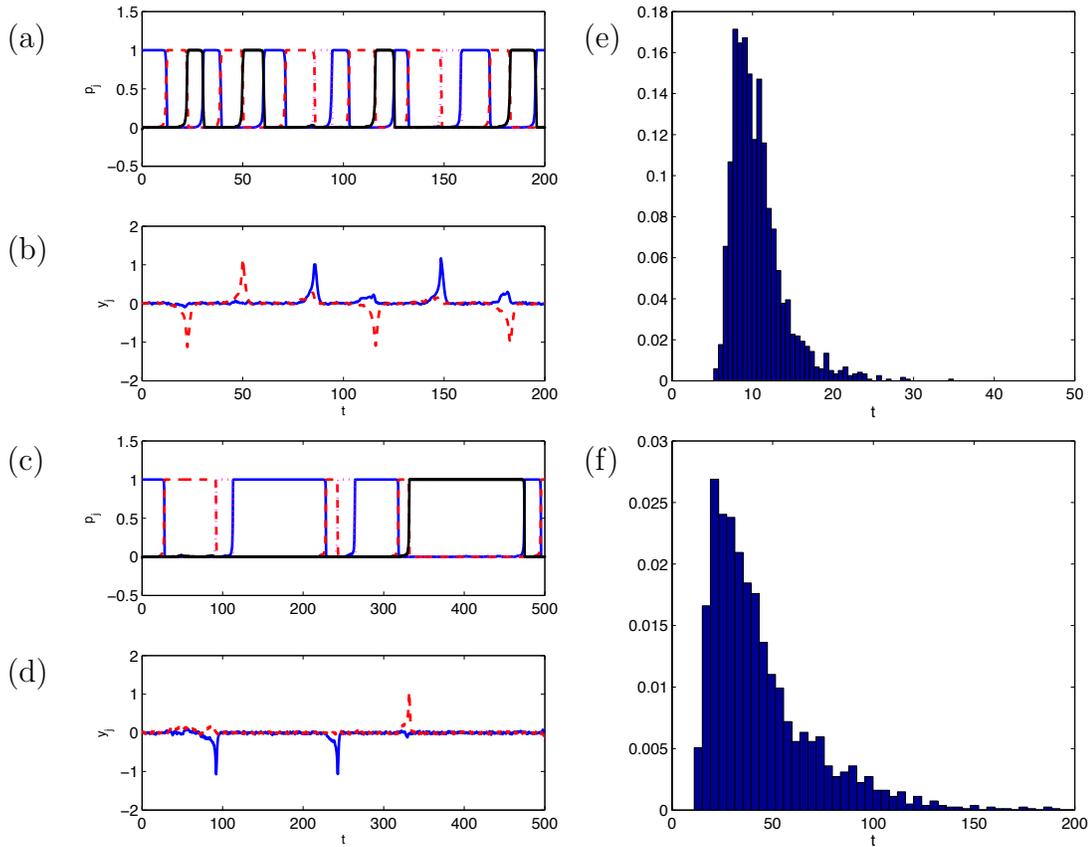

Let us consider a specific example of a realization of the network shown in Figure~\ref{fig:KS}(a) using the coupled cell network illustrated in (b). To this end we consider the model perturbed by additive noise $w_i$:
\begin{equation}
\begin{split}
\dot{p}_1&=p_1(F(1-p^2)+D(p_1^2p^2-p^4))+E(-y_1^2 p_1 p_2+y_4^2p_3^2+y_5^2 p_4^2)+\eta_p w_1\\
\dot{p}_2&=p_2(F(1-p^2)+D(p_2^2p^2-p^4))+E(-y_2^2 p_2p_3-y_3^2p_2p_4+y_1^2 p_1^2)+\eta_p w_2\\
\dot{p}_3&=p_3(F(1-p^2)+D(p_3^2p^2-p^4))+E(-y_4^2 p_3p_1+y_2^2 p_2^2)+\eta_p w_3\\
\dot{p}_4&=p_4(F(1-p^2)+D(p_4^2p^2-p^4))+E(-y_5^2 p_4p_1+y_3^2 p_2^2)+\eta_p w_4\\
\dot{y}_1&=g(y_1,A-B p_1^2+C(y^2-y_1^2))+\eta_1 w_5\\
\dot{y}_2&=g(y_2,A-B p_2^2+C(y^2-y_2^2))+\eta_2 w_6\\
\dot{y}_3&=g(y_3,A-B p_2^2+C(y^2-y_3^2))+\eta_3 w_7\\
\dot{y}_4&=g(y_4,A-B p_3^2+C(y^2-y_4^2))+\eta_4 w_8\\
\dot{y}_5&=g(y_5,A-B p_4^2+C(y^2-y_5^2))+\eta_5 w_9
\end{split}
\label{eq:kssystem}
\end{equation}
where we choose the parameters as in (\ref{eq:hetsystemparams}) except we allow different noise amplitudes in the $y_i$ directions. Figure~\ref{fig:KS-TS} (a) and (b) show a time series for this case, where the parameters are chosen so that a heteroclinic network exists. Figure~\ref{fig:KS-TS}(e) shows a histogram of the residence times near equilibria for a much longer time series.

Stone and colleagues~\cite{StoArm99,StoHol90} have shown that for a heteroclinic cycle, the mean residence time near equilibria scales like $(1/\lambda) \log(\eta)$ for small noise amplitude $\eta$, where $\lambda$ is the expanding eigenvalue at the equilibrium. We note that for the data shown in Figure~\ref{fig:KS-TS}(e), the residence times (not shown) near equilibrium $\xi_2$ are smaller than near the other three equilibria: this is due to the two possible `escape routes' from that equilibrium. 

\subsection{Bifurcation to an excitable Kirk-Silber network}

If $B>A+1$, the equilibria $\xi_j$ are connected to form an excitable network. We consider this same example (\ref{eq:kssystem}) with parameters as in (\ref{eq:hetsystemparams}) except for $B=1.49$, and $
\eta_j=3\times 10^{-5}$. In this case there is an excitable connection with threshold 
$$
\delta_{th}:\approx \sqrt{1-\sqrt{B-A}} \approx 0.07071
$$
from (\ref{eq:delta}).
Figure~\ref{fig:KS-TS} (c) and (d) show a time series for this case. Figure~\ref{fig:KS-TS}(f) shows a histogram of the residence times near equilibria for a much longer time series. Similarly to in the heteroclinic case, we again note that the residence times near $\xi_2$ appear to be smaller than near the other equilibria due to the presence of two escape routes. We also note that the shape of the distributions of escape times in the heteroclinic and excitable cases appears to be different.

In particular, very long residence times are more likely in the excitable case than in the heteroclinic case (the distribution has a fatter tail). This can be seen in comparison of Figure~\ref{fig:KS-TS}(e) and (f); although the timescales are different (due to the different parameters), we have scaled the $x$-axes so that the mean residence times appear at the same position in each figure (these are approximately $10.3$ and $42.3$ respectively), and the fatter tail in (f) can be clearly seen, along with a much more peaked distribution in (e). 

In contrast to the case for the heteroclinic network, we expect the residence times for the excitable network will be governed by a Kramers-type of escape process giving rise an exponential tail in the residence times. This and other statistical properties of switching and residence times near heteroclinic and excitable networks are being investigated for a forthcoming work~\cite{AshPos15}.

\section{Discussion}
\label{sec:discuss}

The main results of the paper are the constructive model (\ref{eq:realiseode}) and Theorems~\ref{thm:realisehet} and \ref{thm:realiseexc} that give open sets of parameter values where one can robustly realize an arbitrary directed graph as an attracting heteroclinic network or an excitable network in phase space.

Although the noise-free dynamics of the network is to some extent trivial (there will be slowing-down heteroclinic dynamics visiting a sequence of equilibria that depends on the initial condition for the heteroclinic network, or, the trajectory will remain at the first stable equilibrium for the excitable network) the dynamics of the networks become much more interesting in the presence of noise. It is known that addition of low noise to an attracting heteroclinic network can lead to random switching around a heteroclinic network in a temporally fairly regular manner \cite{AshPos13}; similarly, addition of noise to an excitable network leads also to random switching around an excitable network. We will explore elsewhere the switching probabilities and residence times near equilibria as a function of the noise strengths and parameter values \cite{AshPos15}.

Similarly, addition of very small inputs in the form of impulses to the $y_k$ variables allows one to control transitions between states in a way that depend on inputs and current state - and so perform finite state computing in the system (\ref{eq:realiseode}) in a manner similar to \cite{ashwin_borresen_04,AshOroWorTow07,neves_timme_12,wordsworth_ashwin_08}. It will be interesting to explore the computational potential of this network. Our construction is ``wasteful'' in the sense that only one cell will be active at any time; the encoding of states is very sparse compared to what nervous systems presumably achieve. It will be a challenge to see whether this construction can be adapted to achieve more dense encoding without losing the high level of control of the dynamics.

Our concept of an excitable network (in phase space) needs to be distinguished from the more general concept of a network of excitable systems. A network of the latter type may or may not realise the former as a network in phase space, depending the nature and strength of the coupling. For example, networks of excitable systems (see for example \cite{Huett2012,KinouchiCopelli2006}) may have many excitable states corresponding to various combinations of cells being active. 

The boundary between heteroclinic and excitable network dynamics for \eqref{eq:realiseode} is on the line $B=1+A$ and corresponds to a subcritical pitchfork bifurcation of the equilibria $\xi_k$ within the invariant subspace  $P_{\ell}$ for each outgoing direction $y_\ell$ from $\xi_k$. By considering $A_{\ell}$ and $B_{\ell}$ (i.e. $A$, $B$ depending on $\ell$) one can clearly design networks using (\ref{eq:realiseode}) that mix heteroclinic and excitable connections \red{ with thresholds $\delta_{\ell}$ that may vary from one connection to another}.

The presence of microscopic noise in the heteroclinic network will result in trajectories wandering around the embedded graph with random choice of outgoing edges at each node. For $A$ and $B$ independent of $k$ and low amplitude noise, this will appear to be a one-step Markov process with a distribution of residence times. However, varying $A_k$ and $B_k$ in the heteroclinic network case introduces the possibility of ``lift-off'' and ``memory'' of the system trajectories in (\ref{eq:realiseode}) as discussed for a related system in \cite{AshPos13}. If the transition past a node is sufficiently fast compared to previous nodes the transition probabilities between nodes may depend not just on the current state but on previously visited states; this will be dependent on the eigenvalues of the equilibria $\xi_k$. This gives the possibility of designing a system with more complex time correlations than a one-step Markov process.

\subsection*{Acknowledgments}

We thank the following for stimulating conversations that contributed to the development of this paper: Mike Field, Marc Timme, John Terry, Ilze Ziedins. We also thank the London Mathematical Society for support of a visit of CMP to Exeter, the University of Auckland Research Council for supporting a visit of PA to Auckland during the development of this research. We are grateful to the Mathematics Departments at both Exeter and Auckland Universities for their hospitality during these visits.

\bibliographystyle{plain}

\end{document}